\documentclass[aps,preprint]{revtex4}%
\usepackage{amsfonts,amsmath,amssymb}
\usepackage{graphicx}

\usepackage{color}

\begin{document}
\preprint{Manuscript}
\title{{\large \textbf{Kinetic Alfv\'{e}n solitary and rogue waves in superthermal plasmas}}}
\author{A. S. Bains}
\affiliation{Shandong Provincial Key Laboratory of Optical Astronomy \& Solar-Terrestrial
 Environment, School of Space Science and Physics, Shandong University at Weihai,
 264209, Weihai, PR China }
\author{Bo Li}
\email{bbl@sdu.edu.cn}
\affiliation{Shandong Provincial Key Laboratory of Optical Astronomy \& Solar-Terrestrial
 Environment, School of Space Science and Physics, Shandong University at Weihai,
 264209, Weihai, PR China }
\author{Li-Dong Xia}
\affiliation{Shandong Provincial Key Laboratory of Optical Astronomy \& Solar-Terrestrial
 Environment, School of Space Science and Physics, Shandong University at Weihai,
 264209, Weihai, PR China }


\begin{abstract}
We investigate the  small but finite amplitude  solitary Kinetic Alfv\'{e}n waves (KAWs) in low $\beta$ plasmas with superthermal electrons modeled by a kappa-type distribution. A nonlinear Korteweg-de Vries (KdV) equation describing the evolution of KAWs is derived by using the standard reductive perturbation method. Examining the dependence of the nonlinear and dispersion coefficients of the KdV equation on the superthermal parameter $\kappa$, plasma $\beta$ and  obliqueness of propagation, we show that these parameters may change substantially the shape and size of solitary KAW pulses. Only sub-Alfv\'enic, compressive solitons are supported. We then extend the study to examine kinetic Alfv\'en rogue waves by deriving a nonlinear Schr\"{o}dinger equation from {the KdV} equation. Rational solutions that form rogue wave envelopes are obtained. We examine how the behavior of rogue waves depends on the plasma parameters in question, finding that the rogue envelopes are lowered with increasing electron superthermality whereas the opposite is true when the plasma $\beta$ increases. The findings of this study may find applications to low $\beta$ plasmas in astrophysical environments where particles are superthermally distributed.
\end{abstract}
\date{\today}
\maketitle

\section{Introduction}

Kinetic Alfv\'{e}n waves ({KAWs})  play an important role in transporting and dissipating energies in a large number of
   space and heliospheric plasma environments~\cite{hasegawa,cramer}.
Unlike the shear Alfv\'en waves in ideal Magnetohydrodynamics (MHD), KAWs are dispersive,
   and have  an electrostatic field parallel to the external magnetic field.
This distinct behavior takes place when kinetic effects associated with the finite ion Larmor radius and/or finite inertial
   length have to be taken into account.
For instance, when the perpendicular wavelength is comparable to the ion gyroradius,
    ions no longer follow the magnetic lines of force, whereas
    electrons still do due to their small Larmor radii.
This produces a charge separation, and the corresponding coupling of shear Alfv\'en waves to electrostatic modes results in
    KAWs as a normal mode of the system.

Dispersion combined with nonlinearity lead to the formation of solitary KAWs.
In 1976, Hasegawa and Mima~\cite{hasegawa76} were the first to find the exact solitary  KAW solutions using kinetic
   theory for a small but finite plasma beta such that $m_e/m_i \ll \beta \ll 1$, where $m_e/m_i$ denotes the electron to ion mass ratio.
These authors neglected ion inertia in the ion continuity equation and  also the ion contribution
   to the parallel current density in Ampere's law, and
   derived the energy balance equation governing the solitary wave solutions of density humps.
Incorporating these two effects while addressing  the same situation where $m_e/m_i \ll \beta \ll 1$, Yu and Shukla~\cite{yu78} showed that
    an upper limit exists for the amplitude of exact solitary Alfv\'{e}n waves.
Shukla et al.\cite{shukla82} again extended the work of Hasegawa and Mima \cite{hasegawa76} by considering the electron inertia to be the dominant
   contributor to wave dispersion as happens when $\beta \ll m_{e}/m_{i} \ll 1$.
They found super-Alfv\'{e}nic solitons with density depression and also derived the KdV equation from the Sagdeev potential equation.
Examining the same model as Shukla et al.\cite{shukla82} but capitalizing on the concept of ion drift velocity,
   Kalita and Kalita~\cite{kalita86} investigated the existence of exact nonlinear  KAWs
   and showed that both super- and sub- Alfv\'{e}nic rarefactive solitons exist, depending on the angle between
   the direction of propagation and the external magnetic field.
Das et al.~\cite{das89} studied the stability of the  solitary KAWs considered by Shukla et al.~\cite{shukla82}.
After deriving the modified KdV equation and then carrying out a stability analysis using the small-$k$ perturbation expansion
   method of Rowlands and Infeld~\cite{rowland,infeld}, these authors found that the growth rate of the instability decreases with increasing angle between the direction of propagation and the external uniform magnetic field.
Ghosh and Das~\cite{ghosh94} again studied the stability of solitary  KAWs for a plasma with
   low but finite $\beta$ values by considering
   the Boltzmann-distributed electrons.
It was observed that the growth rate of the instability attains a maximum for perturbations given
   along a direction lying in the plane containing the external magnetic field
   and the propagation direction of solitary waves.
Over the last two decades or so these investigations attracted the attention of many researchers to study solitary
   kinetic Alfv\'{e}n waves in a variety of plasma models\cite{wu95,wu96,wu96a,wang98,wang98a, wu04,yang05}.
Some investigations were also focused on the effects of dust particles on solitary KAWs~\cite{mahmood02,mahmood08,woo10}.
We note that in most of the above-mentioned theoretical investigations the particle distributions are assumed to be a Maxwellian.

There are many circumstances in which the well-known Maxwellian distribution
     is not a proper description
    of the plasma species \cite{christon,maksimovic, leubner04}.
On theoretical grounds, \cite{hasegawa85} showed that a plasma in the presence of superthermal particles is subject to velocity-space diffusion,
    which will lead to a power law distribution at speeds much higher than the electron thermal speed.
Observationally speaking, the $\kappa$ distribution~\cite{vasy} has been suggested to be more appropriate
    for describing most space plasmas than a Maxwellian.
The presence of a high-energy tail component in a kappa distribution considerably changes the rate of resonant energy transfer between particles
    and plasma waves,  so the conditions for various plasma instabilities may differ substantially for the two distributions.
While there are many studies on the linear and nonlinear characteristics of electrostatic waves using these types
    of distributions \cite{hellberg02,balaku08,saini09,bains10,bains10a},
    only a few exist that examine solitary  KAWs in the presence
    of non-thermally distributed electrons~\cite{bandyo00,roychoudhury,gogoi10,liu11}.
Bandyopadhyay and Das \cite{bandyo00} studied the stability of  KAWs and ion-acoustic waves in nonthermal plasmas using
    the Rowlands-Infeld method.
Solitary  KAWs of arbitrary amplitude for systems where electrons are nonthermally distributed
    is studied by Roychoudhury~\cite{roychoudhury}.
An exact Sagdeev potential equation was derived and it was found that both hump and dip solitons exist
    depending upon the nonthermal parameters for electrons.
Recently, Gogoi and Khan \cite{gogoi10} studied the arbitrary amplitude kinetic Alfv\'{e}n solitons using warm
    adiabatic ions and kappa-distributed electrons in a magnetized plasma.
The system was shown to support compressive and double layer solutions, and  it was demonstrated that the spectral
    index $\kappa$ has significant effects on the shape and size of the solitary KAWs.

Recently rogue waves have become of considerable interest in the broad areas of
    nonlinear fibre optics~\cite{kibler}, optical systems~\cite{solli,hohmann}, atmospheric research~\cite{stenflo} and
    plasmonics~\cite{moslam11}.
Rogue waves are short-lived phenomena appearing suddenly out of normal waves.
Excess amount of energy accumulated in a small region makes the oscillations much stronger than the surrounding waves, with the initial
    process forming rogue waves usually attributed to the modulational instability.
In the context of plasma physics, rogue waves have been studied for both electrostatic~\cite{abdel11,awady11,sabry12} and
    electromagnetic~\cite{shukla12,anuraj13,awady14} modes.
Shukla and Moslem \cite{shukla12} studied the formation of  left- and right-hand  circularly polarized Alfv\'enic rogue waves
    due to the nonlinear interaction between  circularly polarized dispersive Alfv\'en waves and low-frequency electrostatic perturbations.
The amplitude of Alfv\'enic rogue waves decreases with  increasing plasma number density,
    and increases with increasing magnetic field strength.  Panwar et al.\cite{anuraj13} studied the Alfv\'en rogue and solitary waves in an MHD plasma.
It was shown that compressional solitons exist, and the amplitude of the solitons increases with increasing soliton speed  and
    decreases with the plasma $\beta$.
In addition, the envelope of the compressional Alfv\'en rogue waves decreases with increasing $\beta$,
    while electron inertia supports compressional Alfv\'en rogue waves with larger amplitudes. 
\textbf{Very recently, El-Awady et al.\cite{awady14} examined magnetosonic rogue waves in a two-component plasma, 
    and found that a stronger magnetic field leads to a smaller rouge wave amplitude.}
These investigations stimulate our interest in studying rogue wave solutions in the context of KAWs.

The aim of the present manuscript is to elucidate the effects of electron superthermality, as manifested through the commonly observed kappa distribution, on the propagation characteristics of nonlinear kinetic Alfv\'en solitons. An exact KdV equation for the
solitary KAWs is derived with the reductive perturbation method for the first time. The results agree closely with what was found with the well-known Sagdeev potential approach. A nonlinear Schr\"{o}dinger wave equation is then derived to study kinetic Alfv\'en rogue waves.

The organization of this manuscript is as follows.
Section II presents the detailed derivation of the KdV equation, while section III examines the rogue wave solutions.
The last section  summarizes the present study.

\section{Model Equations and Derivation of the KdV equation}
We consider a magnetized electron-ion plasma with small but finite $\beta$ such that $m_e/m_i \ll \beta \ll 1$. Ions are described by a set of fluid moment equations, while electrons are modeled by a kappa velocity distribution. We take the ambient magnetic field ${\bf B}_{0}$ to be in the $z$-direction. The quasi-neutrality condition is used, i.e., $n_e=n_i=n$. The two-potential theory~\cite{kadomtsev} is employed to handle the perpendicular ($\phi_\perp$) and parallel ($\phi_\parallel$) potentials, justifiable for low-$\beta$ plasmas. We assume that the propagation occurs in the $x-z$ plane. The model equations can be written in normalized form as~\cite{yu78}
\begin{align}
\frac{\partial n}{\partial t}+ \frac{\partial (n v_{x})}{\partial{x}}
        + \frac{\partial (n v_{z})}{\partial{z}}=0, \label{1}   \\
\frac{\partial v_z}{\partial t}+ v_{x} \frac{\partial v_{z}}{\partial{x}}
        + v_z \frac{\partial  v_{z}}{\partial{z}}= - \frac{\beta}{2}\frac{\partial \phi_\parallel}{\partial z} , \\
v_x  = - \frac{\beta}{2}\frac{\partial^2 \phi_\perp}{\partial{x} \partial{t}}, \\
\frac{\partial^4 {(\phi_\perp-\phi_\parallel)}}{\partial x^2 \partial z^2}
     =  \frac{2}{\beta}\left[\frac{\partial^2 n}{\partial t^2}+ \frac{\partial (n v_z)}{\partial{t} \partial{z}}\right] , \label{4}
\end{align}
with the electron density given by~\cite{balaku08,saini09}
\begin{equation}
n= \left(1-\frac{\phi_\parallel}{\kappa-3/2}\right)^{-\kappa+1/2}.
\end{equation}
Here time is normalized by the reciprocal of the ion cyclotron frequency $\Omega_{ci}$,
   and the spatial coordinates by the ion inertial length $d_i = c/\omega_{pi}$ where $c$ is  the light speed
   and $\omega_{pi}$ is the ion plasma frequency.
Besides, $n$ is normalized by the ambient density $n_{0}$, the velocity components ($v_{x}, v_{z}$)
   by the Alfv\'en speed $v_{A}= cB_{0}/( n_{0}m_{i}/\epsilon_{0})^{1/2}$,
   and the potentials ($\phi_\perp,\phi_\parallel$) by $T/e$.
Here $T$ is a constant electron temperature, and $e$ is the absolute electron charge.
To derive the KdV equation we use the following stretching coordinates
\begin{equation}
\xi= \epsilon^{1/2} (l_{x}x+l_{z} z-Mt), \quad
\tau=\epsilon^{3/2} t \label{eq_str_coor}
\end{equation}
where $\epsilon$ is a small dimensionless parameter representing the strength of nonlinearity,
   $l_{x}$ and $l_{z}$ are the directional cosines in the $x$- and $z$- directions, respectively ($l_{x}^2+l_{z}^2=1$).
When appropriate, we will also denote the propagation obliqueness with $\theta = \arccos l_z$.
Besides, $M$ is the  wave phase speed in units of $v_{A}$.
The dependent variables can be expanded as (see~\cite{ghosh94,bandyo00})
\begin{align}
& n =  1  +\epsilon n^{(1)}+ \epsilon^{2} n^{(2)}+ ......\nonumber\\
& v_{z} =  \epsilon v_z^{(1)}+ \epsilon^{2} v_z^{(2)}+ ......\nonumber\\
& v_{x} =  \epsilon v_{x}^{(1)} + \epsilon^2 v_{x}^{(2)}+......\nonumber\\
& \phi_\parallel  =  \epsilon \phi_\parallel^{(1)}+ \epsilon^{2} \phi_\parallel^{(2)}+ ......\nonumber\\
& \phi_\perp  =  \phi_\perp^{(1)}+ \epsilon \phi_\perp^{(2)}+ .....
\label{eq_expansion}
\end{align}
 We now plug these expansions into the basic equations,
   and compare terms  at different orders of $\epsilon$.
 At $\epsilon^{3/2}$ and $\epsilon^2$ one finds
\begin{align}
& - c_{1} M \phi_\parallel^{(1)} + l_{x} v_{x}^{(1)}+ l_{z} v_z^{(1)}=0 ,    \label{8} \\
& M v_z^{(1)}= \frac{l_{z}\beta}{2}\phi_\parallel^{(1)} ,  \label{9} \\
& v_{x}^{(1)} = \frac{l_{x} \beta M}{2}\frac{\partial^2 \phi_\perp^{(1)}}{\partial \xi^2} ,  \label{10}  \\
& l_{x}^2 l_{z}^2 \frac{\partial ^2 \phi_\perp^{(1)}}{\partial \xi^2} = \frac{2}{\beta}\left(c_{1} M^2 \phi_\parallel^{(1)}- l_{z} M v_z^{(1)}\right) ,  \label{11}
\end{align}
where $c_{1}= {(\kappa-1/2)}/{(\kappa-3/2)}$.
 Solving Eqs.(\ref{8}) to (\ref{11}) yields the following dispersion relation,
 \begin{equation}
2 c_{1} M^4 -(\beta+2 c_{1})l_{z}^2 M^2+ \beta l_{z}^4=0 , \label{12}
 \end{equation}
    which gives rise to two different modes.
One is a kinetic Alfv\'{e}n mode corresponding to
 \begin{equation}
 M^2 = l_{z}^2,
 \end{equation}
the other corresponds to the well-studied ion acoustic mode~\citep[e.g.,][]{kadijani, sultana} and is not to be pursued here.
 At $\epsilon^{3/2}$, one finds
 \begin{align}
&  c_{1}\frac{\partial \phi_\parallel^{(1)}}{\partial \tau}- c_{1}M \frac{\partial \phi_\parallel^{(2)}}{\partial \xi}
        - c_{2} M \frac{\partial (\phi_\parallel^{(1)})^2}{\partial \xi}
        + l_{x}\frac{\partial v_{x}^{(2)}}{\partial \xi}
        + c_{1} l_{x}\frac{\partial(\phi_\parallel^{(1)} v_{x}^{(1)})}{\partial \xi}+ l_{z} \frac{\partial v_{z}^{(2)}}{\partial \xi}+ c_{1} l_{z} \frac{\partial (\phi_\parallel^{(1)}v_z^{(1)})}{\partial \xi}=0, \\
&  v_{x}^{(2)}
        = \frac{l_{x}\beta M}{2}\frac{\partial^2 \phi_\perp^{(2)}}{\partial \xi^2}
        - \frac{l_{x} \beta}{2}\frac{\partial \phi_\perp^{(1)}}{\partial \xi \partial \tau}, \\
& \frac{\partial v_z^{(1)}}{\partial \tau}- M \frac{\partial v_z^{(2)} }{\partial \xi}
       + l_{x} v_{x}^{(1)}\frac{\partial v_z^{(1)}}{\partial \xi}+ l_{z} v_z^{(1)}\frac{\partial v_z^{(1)}}{\partial \xi}+ \frac{l_{z}\beta}{2}\frac{\partial \phi_\parallel^{(2)}}{\partial \xi}=0 .
 \end{align}
From Eq.(4) at $\epsilon^3$ one finds
\begin{align}
l_{x}^2 l_{z}^2 \frac{\partial^4 (\phi_\perp^{(2)}-\phi_\parallel^{(1)})}{\partial \xi^4}
& = \frac{2}{\beta}\left[c_{1}M^2 \frac{\partial^2 \phi_\parallel^{(2)}}{\partial \xi^2}+ c_{2}M^2 \frac{\partial^2 (\phi_\parallel^{(1)})^2}{\partial \xi^2}
    - 2 c_{1}M \frac{\partial^2 \phi_\parallel^{(1)}}{\partial \tau \partial \xi} \right. \nonumber \\
    &\left.+ l_{z} \frac{\partial v_z^{(1)}}{\partial \xi \partial \tau}- c_{1} l_{z} M \frac{\partial^2 (\phi_\parallel^{(1)} v_z^{(1)})}{\partial \xi^2}- l_{z}M  \frac{\partial^2 v_z^{(2)}}{\partial \xi^2}\right] ,
\end{align}
{where $c_2= (c_1/2) (\kappa+1/2)(\kappa-3/2)$}. With the aid of Eqs.(\ref{8}) to~(\ref{11}), the above equations simplify to
\begin{align}
& a_{2}\frac{\partial \phi_\parallel^{(1)}}{\partial \tau}- c_{1} M \frac{\partial \phi_\parallel^{(2)}}{\partial \xi}
      + a_{3} \phi_\parallel^{(1)} \frac{\partial \phi_\parallel^{(1)}}{\partial \xi}+ l_{z}\frac{\partial v_z^{(2)}}{\partial \xi}
      + \frac{l_{x}^2 \beta M}{2}\frac{{\partial^3 \phi_\perp^{(2)}}}{{\partial \xi^3}}=0,  \label{17} \\
& \frac{l_{z} \beta}{2 M} \frac{\partial \phi_\parallel^{(1)}}{\partial \tau}- M \frac{\partial v_z^{(2)}}{\partial \xi}
      + a_4\phi_\parallel^{(1)}\frac{\partial \phi_\parallel^{(1)}}{\partial \xi}+ \frac{l_{z}\beta}{2}\frac{\partial \phi_\parallel^{(2)}}{\partial \xi}=0,   \label{18} \\
& l_{x}^2  \frac{\partial^3 (\phi_\perp^{(2)}-\phi_\parallel^{(1)})}{\partial \xi^3}
      =  \frac{2 c_1 M^2}{\beta l_{z}^2} \frac{\partial \phi_\parallel^{(2)}}{\partial \xi}+ a_{5} \phi_\parallel^{(1)}\frac{\partial \phi_\parallel^{(1)}}{\partial \xi}+ a_{6} \frac{\partial \phi_\parallel^{(1)}}{\partial \tau}- \frac{2 M}{l_{z}\beta} \frac{\partial v_z^{(2)}}{\partial \xi},    \label{19}
\end{align}
where
\begin{align}
& a_{1} = \frac{2}{l_{x}^2\beta }- \frac{l_{z}^2}{l_{x}^2 M^2},
    \quad a_{2} = 1- \frac{a_{1}l_{x}^2 \beta}{2},
    \quad a_{3}= a_{1}l_{x}^2 \beta M- M+ l_{z}^2 \beta/M ,\nonumber\\
& a_{4} =   \frac{a_{1}l_{x}^2 l_{z} \beta^2}{4}+  \frac{l_{z}^3 \beta^2}{4 M^2},
    \quad a_{5} = \frac{1}{l_{z}^2}\left(\frac{4c_2  M^2}{\beta}-2 c_1 l_{z}^2\right),
    \quad a_{6}= \frac{1}{l_{z}^2}\left(\frac{l_{z}^2}{M}- \frac{4 c_1  M}{\beta}\right) .
\end{align}
Eliminating second order quantities from Eqs.(\ref{17}) to (\ref{19}) with the help of Eq.(\ref{12}), one arrives at the following KdV equation
\begin{equation}
\frac{\partial \psi}{\partial \tau}+ A\psi\frac{\partial \psi}{\partial \xi}+ B\frac{\partial ^3 \psi}{\partial \xi^3}=0 , \label{kdv}
\end{equation}
    where $\psi= \phi_\parallel^{(1)}$. We note that the parallel potential $\phi_\parallel$ in dimensionless form is required to be small for the reductive perturbation approach to be applicable. However, $\phi_\parallel^{(1)}$ is allowed to be of order unity, since it appears immediately after the smallness parameter in the expansion~(\ref{eq_expansion}).
Furthermore, the  nonlinearity and dispersion coefficients, $A$ and $B$, are given by
\begin{align}
& A = - c_{1}l_{z}, \label{A} \\
& B= - \frac{\beta l_{x}^2 l_{z}}{4(c_{1}-\beta/2)} . \label{B}
\end{align}

The steady state solution to the KdV equation can  be obtained by
   transforming $\xi$ and $\tau$ to $\eta =\xi+m_0 \tau$ and $\tau= \tau$, where $m_{0}$ is a constant velocity normalized by $v_{A}$, and by imposing appropriate boundary conditions for localized perturbations, namely $\psi\rightarrow 0$, $\partial \psi/\partial \eta \rightarrow 0$, $\partial^2 \psi/\partial \eta^2 \rightarrow 0$ when $\eta \rightarrow \pm \infty$.
A possible solution is given by
\begin{equation}
\psi = \psi_{m} {\mathrm{sech}}^2(\eta/\Delta) ,
\label{eq_psi1}
\end{equation}
where  the soliton amplitude and width, $\psi_m$ and $\Delta$, are given by
\begin{align}
\psi_{m} =3 m_{0}/A, \mbox{ and } \Delta = \sqrt{4 B/m_{0}} ,
\label{eq_psim_Delta}
\end{align}
respectively.

The characteristics of the solitary KAWs as described by Eq.(\ref{eq_psi1}) can be examined as follows.
First, let us note that with $\beta \ll 1$, both $A$ and $B$ are negative given that $c_1 > 1$.
For $\Delta$ to be meaningful, $m_0$ has to be negative.
Furthermore, if letting $\eta = \epsilon^{1/2}\tilde{\eta}$,  then one
   notices from Eq.(\ref{eq_str_coor}) that $\tilde{\eta} = l_x x + l_z \left[z-\left(1+\epsilon m_0/l_z\right)t \right]$.
A negative $m_0$ therefore renders the parallel propagation speed in dimensionless form, $1+\epsilon m_0/l_z$,  smaller than unity,
   meaning that only sub-Alfv\'enic, compressive ($\psi_m \propto m_0/A >0$) solitons are permitted.
This agrees with the original Hasegawa-Mima discussion~\cite{hasegawa76} (hereafter HM) even though
   a Maxwellian was assumed for the electron distribution therein.
If rewriting $1+\epsilon m_0/l_z$ as $1/K_z$  as defined in HM, then for weakly nonlinear waves $\delta_M = (K_z^2-1) \ll 1$,
    one finds that $m_0 \approx -\delta_M l_z/(2\epsilon)$.
The density perturbation $\tilde{n} \equiv n-1 \approx \epsilon n^{(1)} = \epsilon c_1 \psi^{(1)}$ may then be expressed as
\begin{align}
\tilde{n} = \tilde{n}_M {\mathrm{sech}}^2(\tilde{\eta}/\tilde{\Delta}),
\end{align}
    where
\begin{align}
\tilde{n}_M = \frac{3 c_1}{2}\delta_M, \hspace{20mm}
   \tilde{\Delta} = \sqrt{\frac{3\beta}{1-\beta/(2c_1)} \frac{l_x^2}{\tilde{n}_M}} ,
\end{align}
   which generalizes the original solution (Eq.(12) in HM) to the case where electrons are $\kappa$-distributed.
In the limiting case $\kappa \rightarrow \infty$, one finds that $c_1 = 1$ and Eq.(12) in HM is readily recovered.

\begin{figure}
\includegraphics[width=10cm]{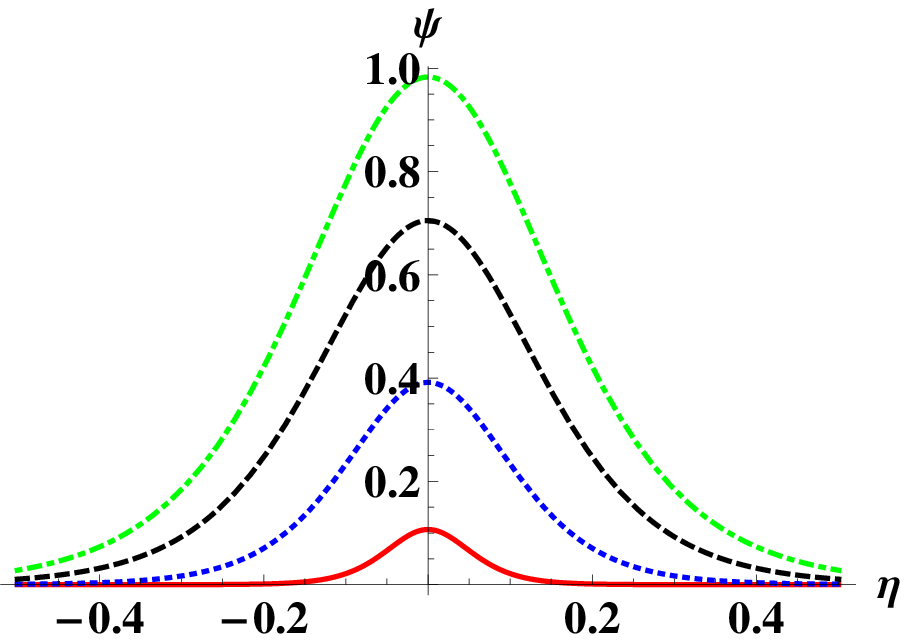}
\caption{(Color online) Profiles of solitary kinetic Alfv\'en waves for a range of $\kappa$ values characterizing the electron
    superthermality
    at fixed values of plasma $\beta = 0.05$ and obliqueness $\theta= 40^\circ$.
    The solid curve corresponds to $\kappa$=1.6, dotted curve to $\kappa$ = 2, dashed curve to $\kappa$=3, and dot-dashed curve to $\kappa$=10. }
\label{fig1}
\end{figure}

As is well-known, the deviation from the Maxwellian behavior of the electron background is
measured by the value of spectral index $\kappa$.
So the effects of the electron superthermality on the soliton amplitude ($\psi_{m}$) and
width ($\Delta$) are investigated in Fig.\ref{fig1},
   which plots the full soliton profile $(\psi)$ as a function of $\eta$ for fixed values of $\beta=0.05$ and $\theta=40^\circ$.
For illustrative purpose, a value of $-0.3$ is adopted for $m_0$.
It can be seen  that with increasing $\kappa$ value (i.e., decreasing superthermality)
    both the amplitude and width of solitary pulses increase,
    a feature readily understandable given that $c_1$ decreases monotonically with increasing $\kappa$.
Moreover,  while the amplitude does not vary significantly with varying $\kappa$ for $\kappa \gtrsim 10$,
   at smaller values of $\kappa$ it is a sensitive function of $\kappa$ with the sensitivity being particularly pronounced when $\kappa$ approaches $3/2$.
 This means that stronger deviations from a Maxwellian support weaker and narrower solitons.

\begin{figure}
\includegraphics[width=10cm]{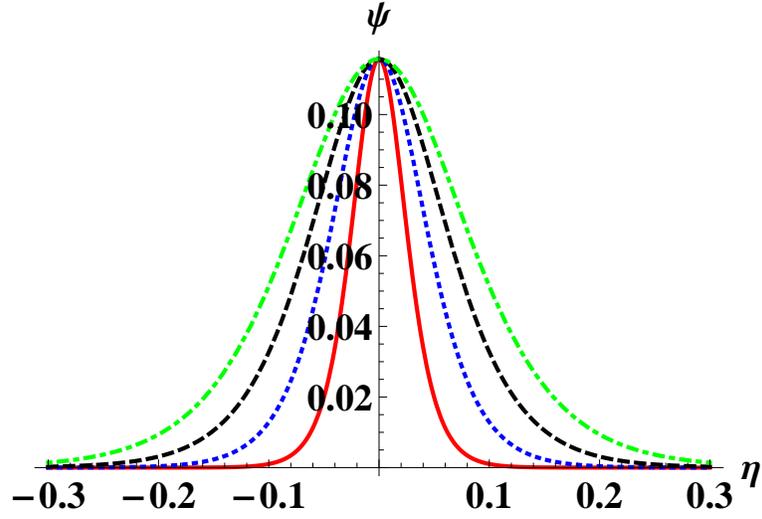}
\caption{(Color online) Profiles of solitary kinetic Alfv\'en waves for a range of values of the plasma $\beta$ at fixed
     values of electron superthermality $\kappa = 2$ and obliqueness $\theta= 45^\circ$.
     The solid curve corresponds to $\beta$=0.01, dotted curve to $\beta$ = 0.03, dashed curve to $\beta$=0.06, and dot-dashed curve to $\beta$=0.1.}
\label{fig2}
\end{figure}

Moving on to Fig.\ref{fig2}, we note that the plasma $\beta$ does not affect the amplitude
   of solitary waves but it has significant effects on their widths, which tend to increase with increasing $\beta$.
Both features are readily understandable with Eq.(\ref{eq_psim_Delta}), which shows that the pulse width $\Delta$
   is approximately proportional to $\sqrt{\beta}$ for small enough $\beta$.
One then sees that plasma systems with high $\beta$ values tend to support  wide solitary KAWs.

\begin{figure}
\includegraphics[width=10cm]{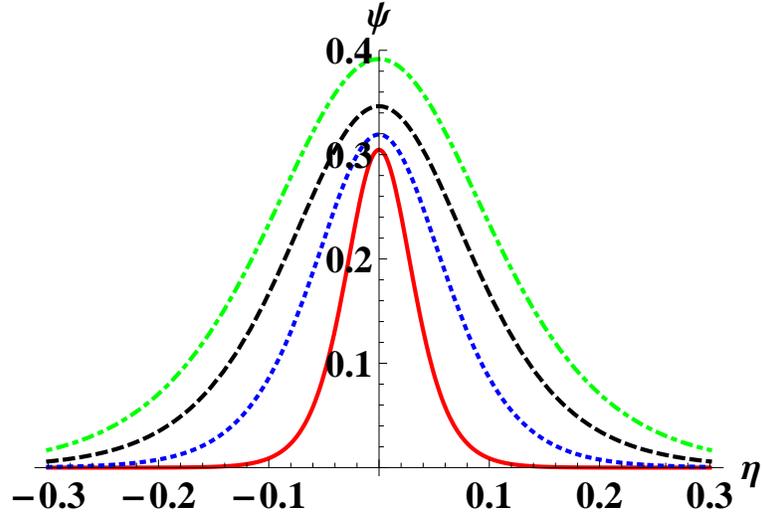}
\caption{(Color online) Profiles of solitary kinetic Alfv\'en waves for a range of values of obliqueness $\theta$ at fixed values of
     electron superthermality $\kappa = 2$ and  plasma $\beta$= 0.05.
     The solid curve corresponds to $\theta=10^\circ$, dotted curve to $\theta = 20^\circ$, dashed curve to $\theta=30^\circ$,
     and dot-dashed curve to $\theta=40^\circ$.}
\label{fig3}
\end{figure}

The effect of propagation angle $\theta$ with respect to the external magnetic field
   is examined in Fig.\ref{fig3}, where the full soliton profile is plotted as a function of $\eta$.
The plasma $\beta$ and $\kappa$ are fixed at $0.05$ and $2$, respectively.
From Fig.\ref{fig3} it can be seen that both the pulse amplitude and width increase
   with increasing  obliqueness, which once again can be understood with Eq.(\ref{eq_psim_Delta}).
We note that the width $\Delta \propto \sqrt{l_x^2 l_z} = \sqrt{\sin^2\theta\cos\theta}$ is zero
   when $\theta$ is either $0^\circ$ or $90^\circ$, meaning that no solitary waves form at exactly parallel or perpendicular propagation.
Furthermore, the width increases when $\theta$ increases from zero, attains its maximum at some angle $\theta \approx 54.7^\circ$, and then
   decreases for further increases in $\theta$.
Considering that the amplitude $\propto 1/\cos\theta$, one sees that the KAW pulses propagating at small angles
   to the external magnetic filed tend to be weak and narrow ones.

\section{Nonlinear Schrodinger equation and Rogue wave solutions}
To study the modulational instability of the weakly nonlinear wave packets described
   by the KdV equation (\ref{kdv}), we first make the dispersion coefficient positive
   by making a simple transform $\overline{\psi}= -\psi$, $\overline{\tau}=\tau$ and $\overline{\xi}=-\xi$,   which will not change the solution in essence \cite{belashov}.
Actually, the KdV equation can be equally applied to a medium with negative dispersion (when the phase velocity of waves decreases with increasing wave number) as well as to a medium with positive dispersion, the difference being
   only in the direction in which the $\xi$- axis is directed.
With this transform the KdV equation retains its original form, but the resulting nonlinearity and dispersion coefficients
   become $\overline{A}=A$ and $\overline{B}=-B$.
Next we consider solutions to the resulting KdV equation in the form of a weakly modulated sinusoidal wave $\overline{\psi}$
    (for details of the technique, see e.g.,\cite{shimizu,labany})
\begin{equation}
\overline{\psi}  =\sum_{n=1}^{\infty}\epsilon^{n}\sum_{l=-\infty}^{+\infty}\psi
_{l}^{n}(\zeta,\chi)e^{\iota l(k \bar{\xi}-\omega \bar{\tau})} \label{30}
\end{equation}
where $k$ is the carrier wave number and $\omega$ is the angular frequency for the given KAW.
The stretched variables $\zeta$ and $\chi$ are chosen as
\begin{equation}
\zeta=\epsilon(\overline{\xi}+v_g \overline{\tau}) \quad \text{and} \quad \chi= \epsilon^2 \overline{\tau} , \label{31}
\end{equation}
where $v_g$ is the group velocity.
By standard practice, we assume that  fast scales enter into our discussion via the phase $(k \xi-\omega \tau)$ only,
   while  slow scales $(\zeta,\chi)$ do so only in the form of the arguments of the $l$-th harmonic amplitude $\psi_{l}^n$.
For $\psi (\zeta,\chi) $ to be real, one must require that $\psi_{-l}^n = \psi_{l}^{n*}$ where $^*$ denotes complex conjugate. Substituting Eqs.(\ref{30}) and (\ref{31}) into Eq.(\ref{kdv}), one readily finds
\begin{eqnarray}
&&-\iota l \omega \psi_{l}^{(n)} + v_{g} \frac{\partial \psi_{l}^{(n-1)}}{\partial \chi}+ \frac{\partial \psi_{l}^{(n-2)}}{\partial \chi} + \overline{A}  \sum_{n'=1}^{\infty} \epsilon^n \sum_{l'= -\infty}^{+\infty} \left(\iota k l \psi_{l}^{(n)}\psi_{l-l'}^{(n-n')} + \psi_{l-l'}^{(n-n')}\frac{\partial \psi_{l}^{(n)}}{\partial \zeta} \right) \nonumber \\
&&+ \overline{B} \left( -\iota l^3 k^3 \psi_{l}^{(n)}- 3 l^2 k^2 \frac{\partial \psi_{l}^{(n-1)}}{\partial \zeta}+ 3 \iota l k \frac{\psi_{l}^{(n-2)}}{\partial \zeta^2}\right) + \overline{B} \frac{\partial \psi_{l}^{(n-3)}}{\partial \zeta^3}=0 .\label{32}
\end{eqnarray}
With $(n, l)=(1, 1)$ one finds that
\begin{equation}
  \omega = - \overline{B} k^3,
\end{equation}
  while the equations with $(n, l)=(2, 1)$ give the group velocity as
\begin{equation}
v_{g} = 3 \overline{B} k^2 .
\end{equation}
Proceeding to $(n, l) = (2, 2)$ and $(2, 0)$, one finds that
\begin{align}
\psi_{2}^{(2)}= \frac{\overline{A}}{6 \overline{B} k^2} \psi_{1}^{(1)}, \hspace{0.2cm}
\psi_{0}^{(2)}= -\frac{\overline{A}}{v_{g}} \left| \psi_{1}^{(1)} \right| .
\end{align}

Finally, substituting the above derived expressions into the $(l=1)$ component
    of the third-order part of the reduced equations, one obtains the following
    nonlinear Schr\"{o}dinger equation expressed in $\Psi\equiv \psi_{1}^{(1)}$,
\begin{equation}
\iota\frac{\partial\Psi}{\partial\chi}+P\frac{\partial^{2}\Psi
}{\partial\zeta^{2}}+Q|\Psi|^{2}\Psi=0, \label{nlse}
\end{equation}
where
\begin{equation}
P =  6 \overline{B} k, \quad {\rm and} \quad Q =  \frac{\overline{A}^2}{6\overline{B}k}
\end{equation}
are the dispersion and nonlinear coefficients, respectively.
To study rogue waves, a rational solution to Eq.(\ref{nlse}) reads (see e.g.,\cite{moslam11a})
\begin{equation}
\Psi = \frac{1}{\sqrt{Q}}\left[\frac{4(1+2 \iota \chi)}{1+ 4 \chi^2 + 4 \zeta^2 /P}\right] \exp{\iota \chi} , \label{rg}
\end{equation}
   which predicts the concentration of KAW energy into a small region due to the nonlinear properties of the plasma medium.
Rogue waves are usually an envelope of a carrier wave with a wavelength smaller than in the central region of the envelope.

Based on the linear stability analysis \cite{amin,bains11}, it is observed that the waves as described by the nonlinear Schrodinger equation
    are modulationally unstable when $P/Q>0$ and when the modulation wave number satisfies
    $K^{2}< |\psi_{0}|^{2} {(2Q)}/{P}$, where $\psi_{0}$ is the amplitude of the carrier waves.
Furthermore, the maximum growth rate is given by $Q|\psi_{0}|^{2}$ and is attained at $K=\sqrt{\frac{Q}{P}}|\psi_{0}|$.
Two types of stationary solutions are possible: (i) Unstable solutions called bright envelope solitons when $P/Q>0$ and
   (ii) Stable solutions called dark envelope solitons when $P/Q<0$.
Evidently, the case examined in the present study falls in the first category.

\begin{figure}[htp]
\centering
 \begin{tabular}{cc}
    \includegraphics[width=80mm]{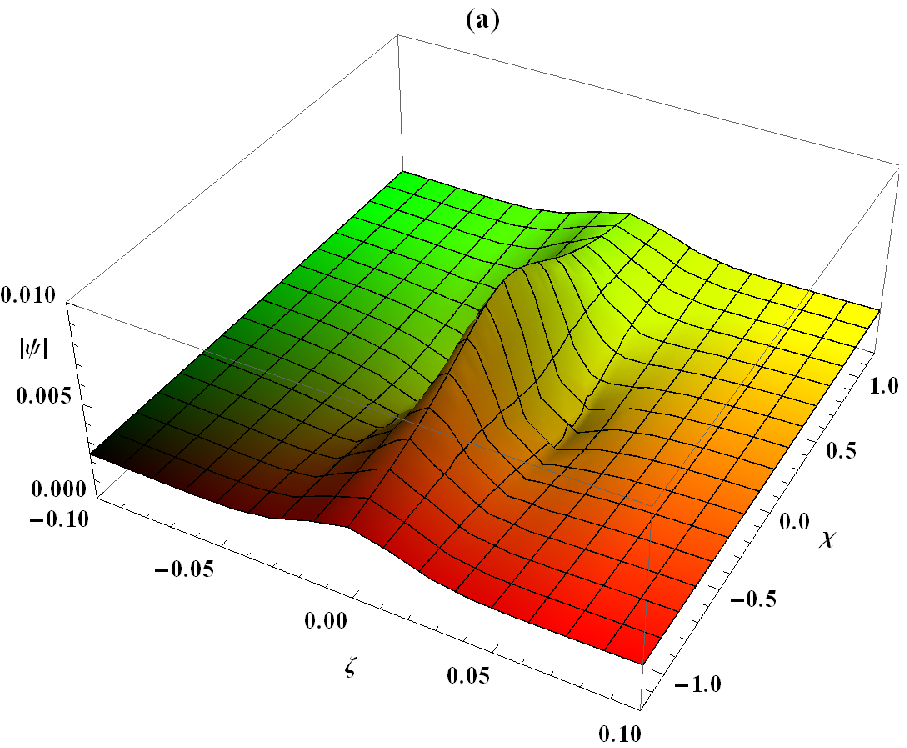} &
    \includegraphics[width=80mm]{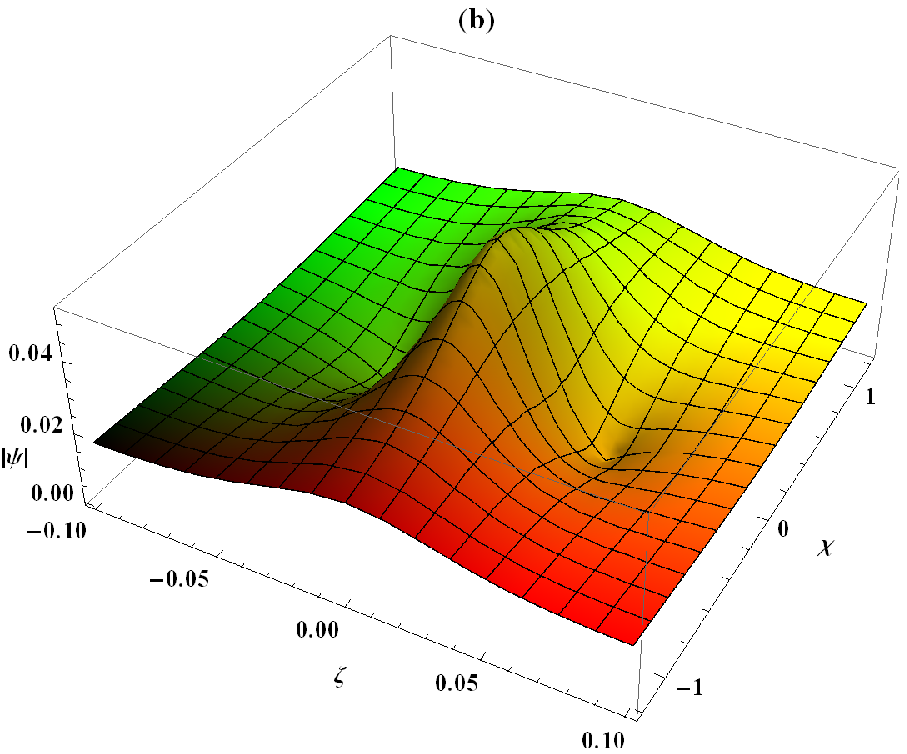} \\
    \includegraphics[width=80mm]{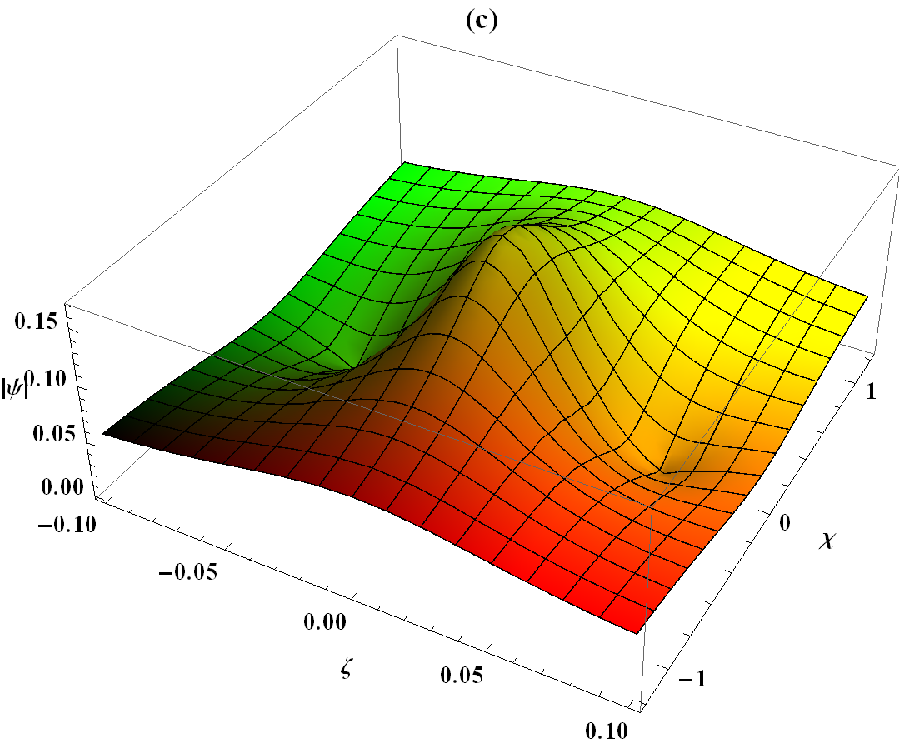}
  \end{tabular}
\caption{(Color online) The rogue wave profile $\left|\Psi\right|$ with respect to $\chi$ and $\zeta$ for different values of $\kappa$:
     (a) $\kappa$ = 1.6, (b) $\kappa$ = 2 and (c) $\kappa$ = 4. Here $\beta$ = 0.05, $\theta$ = 20 and $k = 0.9$.}
\label{fig4}
\end{figure}

To show how the wave envelope depends on the  electron superthermality,
    Figure~\ref{fig4} shows the $\Psi$ profiles
    as a function of $\zeta$ and $\chi$ for (a) $\kappa= 1.6$, (b) $\kappa=2$ and (c) $\kappa=4$
    when $\beta= 0.05, \theta=20^\circ,$ and $k= 0.9$.
It is clear that both the amplitude and width of the envelope increase when $\kappa$ increases,
     meaning that the more pronounced the deviation from a Maxwellian, the weaker the rogue waves.
Thus one can say that rogue waves for Maxwellian plasmas correspond to higher energy
     than for superthermal plasmas.

\begin{figure}[htp]
\centering
 \begin{tabular}{cc}
    \includegraphics[width=80mm]{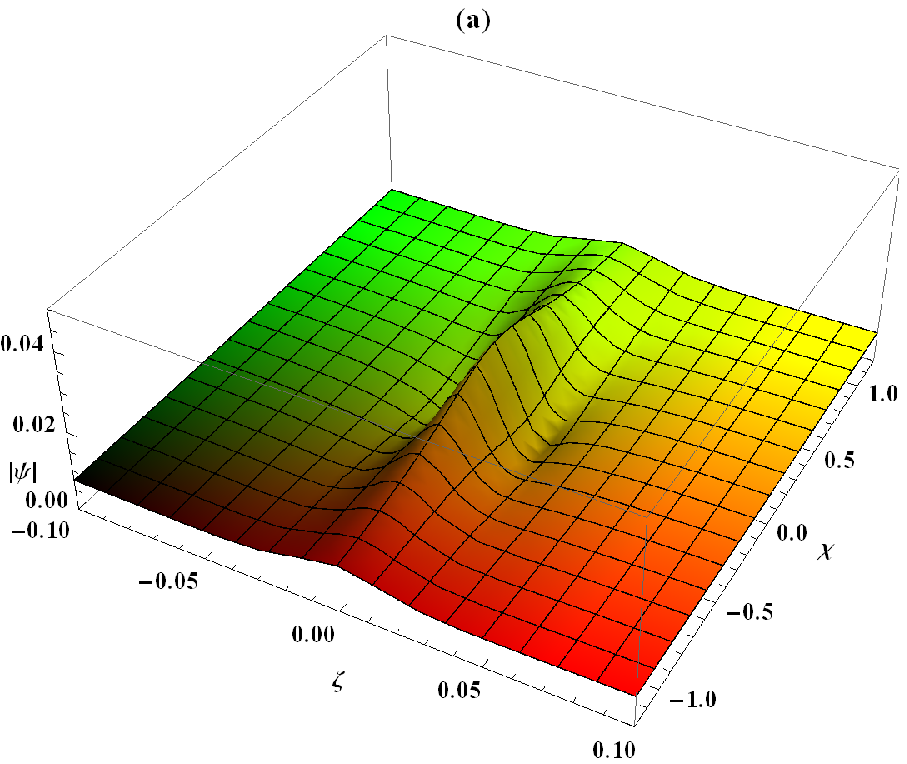} &
    \includegraphics[width=80mm]{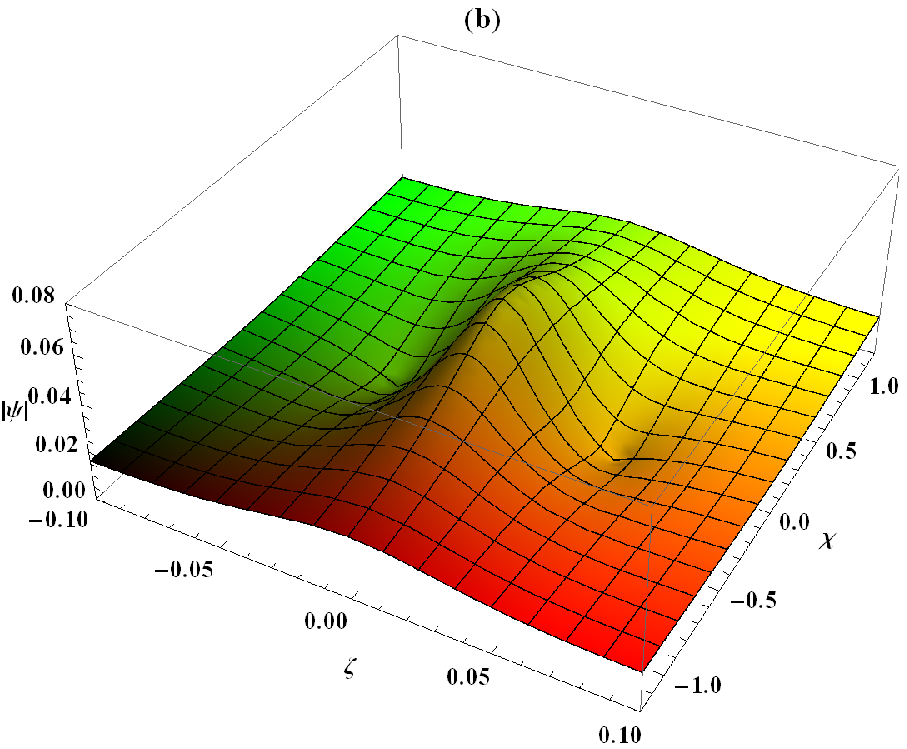} \\
    \includegraphics[width=80mm]{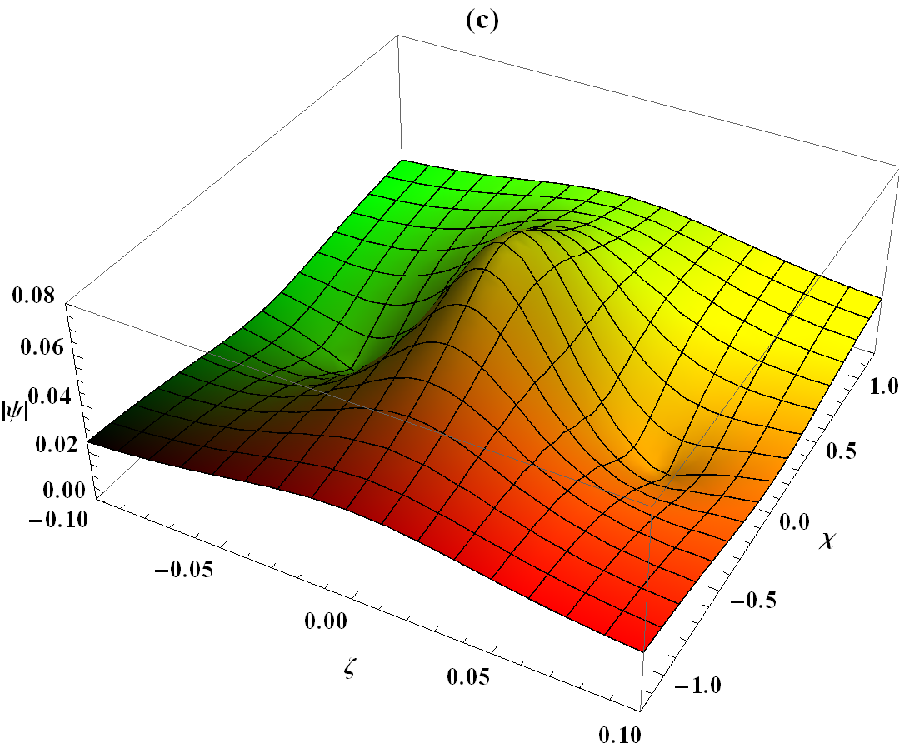}
  \end{tabular}
\caption{(Color online) The rogue wave profile $\left|\Psi\right|$ with respect to $\chi$ and $\zeta$ for different values of $\beta$:
    (a) $\beta$ = 0.01, (b) $\beta$ = 0.05 and (c) $\beta$ = 0.1. Here $\kappa = 2$, $\theta = 20^\circ$ and $k = 0.9$. }
\label{fig5}
\end{figure}

Figure \ref{fig5} examines how the rogue waves depend on the plasma $\beta$ by presenting the
    $\Psi$ profiles as a function of $\zeta$ and $\chi$ for (a) $\beta=0.01$, (b)  $\beta =0.05$ and (c) $\beta = 0.1$
    when $\kappa= 2, \theta =  20^\circ$ and $k= 0.9$.
One can readily see that rogue waves have a sensitive dependence on the plasma $\beta$ for the parameters chosen here,
    with the tendency being that increasing $\beta$ substantially increases the amplitude of the envelope.
Hence one expects to see stronger and narrower envelops in the regions with higher $\beta$.
Although not shown,  the amplitude of rogue waves is found to increase with increasing obliqueness.

\section{Conclusions}
We have presented a theoretical study on the propagation dynamics of solitary kinetic Alfv\'{e}n waves (KAWs)   in low beta plasmas characterized by a superthermally distributed electron population, modeled by a kappa-type distribution. The standard reductive perturbation method was employed to derive the Korteweg de-Vries (KdV) equation. We have traced the effects of the
electron superthermality, plasma $\beta$ and obliqueness on the characteristics of solitary KAWs.

\textbf{
An exact KdV equation has been derived for the first time using the reductive perturbation method for kinetic Alfv\'en solitary waves. 
Only compressive solitons are permitted.
The Hasegawa-Mima~\cite{hasegawa76} results are recovered when
   the electron distribution approaches a Maxwellian.
The electron superthermality makes solitary pulses narrower and weaker, 
   relative to a Maxwellian plasma. 
A magnetic field makes solitons wider but has no effect on the pulse amplitude.
With increasing obliqueness relative
   to the external magnetic field, solitons become taller and wider. 
When the wave vector is exactly parallel to the external magnetic field, no soliton forms, for in this case
   the basic equations takes the form of ideal MHD, which is non-dispersive in nature for uniform plasmas.
To study rogue wave solutions for KAWs, a nonlinear Schr\"{o}dinger wave equation was derived from the KdV equation by 
   using a standard perturbation method. 
It is observed that the electron superthermality lowers the rogue wave amplitude, meaning that kinetic Alfv\'en rogue waves
   can have higher energies in Maxwellian plasmas than in superthermal plasmas. 
Furthermore, it is observed that plasmas with higher $\beta$ support stronger rogue waves.}

Our findings may help explain and interpret nonlinear oscillations in low $\beta$ plasmas such as planetary magnetospheres and the solar wind.

\acknowledgements
We are grateful to the anonymous referee whose comments helped improve this manuscript substantially.
This research is supported by the 973 program 2012CB825601, the National Natural
Science Foundation of China (40904047, 41174154, 41274176, and 41274178), the Ministry of Education
of China (20110131110058 and NCET-11-0305), and by the Provincial Natural Science Foundation of Shandong
via Grant JQ201212.


\clearpage

\clearpage

\clearpage

\clearpage

\clearpage


\begin{thebibliography}{99}

\bibitem{hasegawa} A. Hasegawa and C. Uberoi, The Alfv\'en Wave (Technical Information Center, U. S. Department of Commerce, Springfield, Virginia, 1982).
\bibitem{cramer} N. F. Cramer, The Physics of Alfv\'en Waves, Wiley-VCH, Berlin, 2001.
\bibitem{belcher} J. W. Belcher and L. Davis, Jr., J. Geophys. Res. 76, 3534 (1971)
\bibitem{sagdeev} R. Z. Sagdeev and A. A. Galeev, Nonlinear Plasma Theory (Benjamin, New York, 1969), p. 8.
\bibitem{louarn}P. Louarn, J. E. Wahlund, T. Chust, H de Feraudy, A. Roux, B. Holback, P. O. Dovner, A. I. Eriksson, and G. Holmgren, Geophys. Res. Lett. 21, 1847 (1994).
\bibitem{dovner} P. O. Dovner and G. Holmgreen, Geophys. Res. Lett. 21, 1827 (1994).
\bibitem{hasegawa76} A. Hasegawa and K. Mima, Phys. Rev. Lett. 37, 690 (1976).
\bibitem{yu78} M. Y. Yu and P. K. Shukla, Phys. Fluids 21, 1457 (1978).
\bibitem{shukla82} P. K. Shukla, H. U. Rahaman and R. P. Sharma, J. Plasma Phys. 28, 125 (1982).
\bibitem{kalita86} M. K. Kalita and B. C. Kalita, J. Plasma Phys. 35, 267 (1986).
\bibitem{das89} K. P. Das, L. P. J. Kamp and F. W. Sluijter, J. Plasma Phys. 41, 171 (1989).
\bibitem{rowland} G. Rowlands, J. Plasma Phys. 3, 567 (1969)
\bibitem{infeld} E. Infeld and G. Rowlands, J. Plasma Phys. 10, 293 (1973).
\bibitem{ghosh94} G. Ghosh and K. P. Das, J. Plasma Phys. 51, 95 (1994).
\bibitem{wu95} D. J. Wu, D. Y. Wang and C. G. Falthammar, Phys. Plasmas 2, 4476 (1995).
\bibitem{wu96} D. J. Wu and D. Y. Wang, Phys. Plasmas 3, 4304 (1996).
\bibitem{wu96a} D. J. Wu, G. L. Huang and D. Y. Wang, Phys. Plasmas 3, 2879 (1996).
\bibitem{wang98} X. Y. Wang, X. Y. Wang, Z. X. Liu and Z. Y. Li, Phys. Plasmas 5, 3477 (1998).
\bibitem{wang98a} X. Y. Wang, X. Y. Wang, Z. X. Liu and Z. Y. Li, Phys. Plasmas 5, 4395 (1998).
\bibitem{wu04} D. J. Wu and J. K. Chao,  Nonlin. Processes Geophys. 11, 631 (2004).
\bibitem{yang05} L. Yang and D. J. Wu, Phys. Plasmas 12, 112901 (2005).
\bibitem{mahmood02} M. A. Mahmood, A. M. Mirza, P. H. Sakanaka and G. Murtaza, Phys. Plasmas 9, 3794 (2002).
 \bibitem{mahmood08} S. Mahmood and H. Saleem, Phys. Plasmas 15, 114504 (2008).
 \bibitem{woo10} M. H. Woo, C.M. Ryu and C. R. Choi, Phys. Plasmas 17, 053707 (2010).
 \bibitem {christon}S. P. Christon, D. G. Mitchell, D. J. Williams, L. A.Frank, C. Y. Huang, and T. E. Eastman, J. Geophys. Res. 93, 2562, (1988).

\bibitem {maksimovic} M. Maksimovic, V. Pierrard, and P. Riley, Geophys. Res. Lett. 24, 1511, (1997).

\bibitem {leubner04}M. P. Leubner, Phys. Plasmas 11, 1308 (2004).
\bibitem{hasegawa85} A. Hasegawa, K. Mima, and M. Duong-van, Phys. Rev. Lett. 54, 2608 (1985).
\bibitem{vasy} V. M. Vasyliunas, J. Geophys. Res. 73, 2839 (1968).

 \bibitem{hellberg02} M. A. Hellberg and R. L. Mace, Phys. Plasmas 9, 1495 (2002).
 \bibitem{balaku08} T. K. Baluku and M. A. Hellberg, Phys. Plasmas 15, 123705 (2008).
 \bibitem{saini09} N. S. Saini, I. Kourakis and M. A. Hellberg, Phys. Plasmas 16, 062903 (2009).
 \bibitem{bains10} T. S. Gill, A. S. Bains and C. Bedi, Phys. Plasmas 17, 013701 (2010).
 \bibitem{bains10a} T. S. Gill, A. S. Bains, N. S. Saini and C. Bedi, Phys. Letts. A 374, 3210 (2010).
\bibitem{bandyo00} A. Bandyopadhyay and K. P. Das, Phys. Plasmas 7, 3227 (2000).
\bibitem{roychoudhury} R. Roychoudhury, J. Plasma Phys. 67, 199 (2002).
 \bibitem{gogoi10} R. Gogoi and M. Khan, Phys. Plasmas 17, 112311 (2010).
 \bibitem{liu11} Y. Liu, S. Q. Liu and B. Dai, Phys. Plasmas 18, 092309 (2011).
\bibitem{kibler} B. Kibler, J. Fatome, C. Finot, G. Millot, F. Dias, G. Genty, N.
Akhmediev, and J. M. Dudley, Nat. Phys. 6, 790 (2010).
\bibitem{solli}D. R. Solli, C. Ropers, P. Koonath, and B. Jalali, Nature 450, 1054 (2007).
\bibitem{hohmann} R. Hohmann, U. Kuhl, H.-J. Stockmann, L. Kaplan, and E. J. Heller, Phys.Rev. Lett. 104, 093901 (2010).
\bibitem{stenflo} L. Stenflo and M. Marklund, J. Plasma Phys. 76, 293–295 (2010).
\bibitem{moslam11} W. M. Moslem, P. K. Shukla, and B. Eliasson, EPL 96, 25002 (2011).
\bibitem{abdel11} U. M. Abdelsalam, W. M. Moslam, A. H. Khater and P. K. Shukla, Phys. Plasmas 18, 092305 (2011).
\bibitem{awady11} E. I. El-Awady and W. M. Moslem, Phys. Plasmas 18, 082306 (2011).
\bibitem{sabry12} R. Sabry, W. M. Moslem and P. K. Shukla, Phys. Plasmas 19, 122903 (2012).
\bibitem{shukla12} P. K. Shukla and W. M. Moslem, Phys. Letts. A 376, 1125 (2012).
\bibitem{anuraj13} A. Panwar, H. Rizvi and C. M. Ryu, Phys. Plasmas 20, 082101 (2013).
\bibitem{awady14} E.~I.~El-Awady, H.~Rizvi, W.~M.~Moslem, S.~K.~El-Labany, A. Raouf and M. Djebli, Astrophys. Space Sci. 349, 5 (2014).
\bibitem{kadomtsev} B. B. Kadomtsev, Plasma Turbulence (Academic, New York, 1965), p.82.
\bibitem{kadijani} M. N. Kadijani, H. Abbasi and H. H. Pajouh, Plasma Phys. Control. Fusion 53, 025004 (2011).
\bibitem{sultana} S. Sultana, I. Kourakis, N. S. Saini and M. Hellberg, Phys. Plasmas 17, 032310 (2010).
\bibitem{belashov} V. Yu. Belashov and S. V. Vladimirov,  Solitary waves in Dispersive Complex Media (Springer-Verlag Berlin Heidelberg 2005) p. 22.
\bibitem{shimizu} K. Shimizu and Y. H. Ichikawa, J. Phys. Soc. Jpn 33, 789 (1972).
\bibitem{labany} S. K. El-Labany, J. Plasma Phys. 54, 295 (1995).
\bibitem{moslam11a} W. M. Moslam, Phys. Plasmas 18, 032301 (2011).
\bibitem{amin} M. R. Amin, G. E. Morfill, and P. K. Shukla, Phys. Rev. E 58, 6517(1998).
\bibitem{bains11} A. S. Bains, M. Tribeche, and T. S. Gill, Phys. Lett. A 375, 2059–2063 (2011).
\end{thebibliography}
\end{document}